\documentclass[twocolumn,floatfix,aps,prl,showpacs,superscriptaddress,10pt]{revtex4-2}
\usepackage[utf8]{inputenc}
\usepackage{graphicx}  
\usepackage{dcolumn}   
\usepackage{bm}        
\usepackage{amssymb}   
\usepackage{amsmath}
\usepackage{color}
\usepackage{bbm}

\begin{document}

\title{Classical Spin Liquid State in the $S=\frac{5}{2}$ Heisenberg Kagom\'e Antiferromagnet Li$_9$Fe$_3$(P$_2$O$_7$)$_3$(PO$_4$)$_2$}

\author{E. Kermarrec}
\author{R. Kumar}
\author{G. Bernard}
\author{R. H\'enaff}
\author{P. Mendels}
\author{F. Bert}
\affiliation{Universit\'e Paris-Saclay, CNRS, Laboratoire de Physique des Solides, 91405, Orsay, France}
\author{P. L. Paulose}
\affiliation{Department of Condensed Matter Physics and Materials Science, TIFR, Mumbai}
\author{B. K. Hazra} 
\affiliation{School of Physics, University of Hyderabad, Hyderabad-500046, India}
\author{B. Koteswararao}
\affiliation{Department of Physics, Indian Institute of Technology Tirupati, Tirupati 517506, India}


\begin{abstract}
We investigate the low temperature magnetic properties of a $S=\frac{5}{2}$ Heisenberg kagom\'e antiferromagnet, the layered monodiphosphate Li$_9$Fe$_3$(P$_2$O$_7$)$_3$(PO$_4$)$_2$, using magnetization measurements and $^{31}$P nuclear magnetic resonance. An antiferromagnetic-type order sets in at $T_{\rm N}=1.3$~K and a characteristic magnetization plateau is observed at 1/3 of the saturation magnetization below $T^* \sim 5$~K. A moderate $^{31}$P NMR line broadening reveals the development of anisotropic short-range correlations concomitantly with a gapless spin-lattice relaxation time $T_1 \sim k_B T / \hbar S$, which may point to the presence of a semiclassical nematic spin liquid state predicted for the Heisenberg kagom\'e antiferromagnetic model or to the persistence of the zero energy modes of the kagom\'e lattice under large magnetic fields.
\end{abstract}

\pacs{
75.10.Jm,     
76.60.-k,     
75.30.Cr,     
75.10.Kt,     
75.40.Gb,     
}

\maketitle

Magnetic frustation sets as a playground for the discovery of new exotic phases of quantum matter. The Heisenberg antiferromagnetic model on the kagom\'e lattice (KHAF) constitutes the archetype of the frustration in two dimensions, whose ground-state and exotic spin dynamics still remain topical questions in quantum magnetism even after 30 years of research \cite{IFM,Norman2016,Broholm2020,Khuntia2020,Wang2021}.

\begin{figure}[h]
\includegraphics[width=\columnwidth]{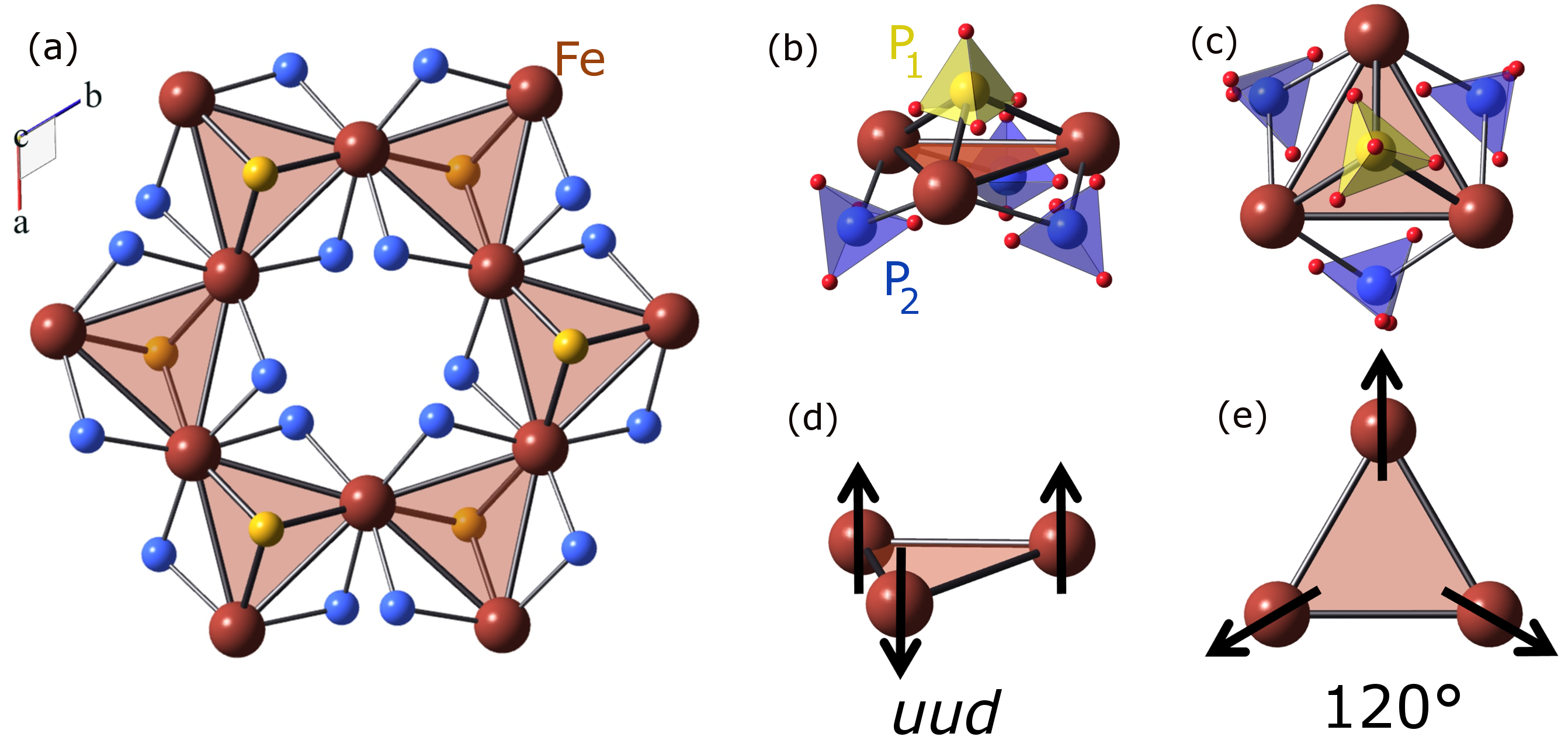}
\caption{\label{kagome} (Color online) (a) Schematic structure of LiFePO illustrating the corner-sharing triangular lattice made of Fe$^{3+}$, $S=\frac{5}{2}$  (brown). Li and O atoms are omitted for clarity. (b), (c) Fe atoms are magnetically coupled through PO$_4$ chemical groups with two different phosphorus sites: P$_1$ (yellow) and P$_2$ (blue). Illustration of the up-up-down \textit{uud} (d) and of the 120$^{\circ}$ ordered structure (e).}
\end{figure}

Early on, Chalker \textit{et al.}\cite{Chalker1992} highlighted the distinctive magnetic properties of the classical ($S=\infty$) KHAF, for which thermal fluctuations select a spin nematic, or coplanar, ground state via the order-by-disorder mechanism in zero field \cite{Villain1980,Henley1989},  with spins  on a single triangle oriented at 120$^{\circ}$ from each other (Fig.\ref{kagome}e). Its spin dynamics is governed by zero-energy soft modes, i.e. special lines of weathervane defects \cite{Ritchey1993}. The order-by-disorder mechanism is also responsible for the remarkable existence of semiclassical fractional plateaus of magnetization in frustrated magnets \cite{IFM,Honecker2004,Kageyama1999,Matsuda2013,Capponi2013,Nishimoto2013}. In particular, the 1/3 plateau --a hallmark of frustration\cite{IFM,Cabra2002}--, arising from collinear \textit{uud} spin configurations (Fig.\ref{kagome}d), has been observed in the triangular compounds Cs$_2$CuBr$_4$\cite{Ono2003} and RbFe(MoO$_4$)$_2$\cite{Inami1996,Svistov2003,Svistov2006} but remains debated and/or difficult to observe for quantum kagom\'e antiferromagnets with large interaction \cite{Okuma2020} or with deviations from the pure KHAF model \cite{Okamoto2011,Ishikawa2015}.

More recently, the phase diagram of the KHAF in an external field was revisited with state-of-the art numerical methods, notably tensor network formalism\cite{Picot2015,Picot2016}, DMRG\cite{Nishimoto2013} or exact diagonalization\cite{Capponi2013}. It encompasses fully ``quantum" phases and semiclassical plateaus, including the 1/3 plateau that appears for all values of spin $S$. The specific aspect of the 1/3 plateau of the classical kagom\'e lattice under an applied field was pinpointed by finite-temperature Monte-Carlo studies which showed that thermal fluctuations favor a classical \textit{collinear spin liquid} phase \cite{Zhitomirsky2002,Gvozdikova2011} at $M=\frac{1}{3}M_{\rm sat}$, made of a macroscopic number of degenerate \textit{uud} states.

The comparison with classical experimental candidates has remained challenging for long as naturally occuring kagom\'e lattice are rare. The jarosites family \textit{A}Fe$_3$(OH)$_6$(SO$_4$)$_2$ (with A$^+$ typically an univalent cation) is held up as a paragon of the classical KHAF\cite{IFM,Greedan2001,Wills2000}. They all (but A$^+$=D$_3$O$^+$\cite{FakEPL2008}) display long range ordering into a $q=0$ structure, as a result of a deficient kagom\'e network and/or of perturbative anisotropic terms. Furthermore, their large antiferromagnetic interaction $J\sim600$~K requires unattainable applied magnetic fields in order to reach the 1/3 plateau. 

In this Letter, we investigate the low temperature magnetic properties of the layered iron monodiphosphate Li$_9$Fe$_3$(P$_2$O$_7$)$_3$(PO$_4$)$_2$ (LiFePO), a $S=\frac{5}{2}$ Heisenberg kagom\'e antiferromagnet material with a moderate exchange $J \sim 1$~K. LiFePO thus offers the opportunity to probe and finely explore experimentally the phase diagram of the classical KHAF under applied field. Our measurements reveal a clear 1/3 magnetization plateau with an extended spin-liquid regime characterized by exotic spin dynamics.

LiFePO belongs to the family of monodiphosphates, which were primarily studied for their electronic transport properties in the context of battery materials \cite{Poisson1998}. LiFePO crystallizes in the centrosymmetric trigonal space group P$\overline{3}c1$ where Fe$^{3+}$ ($S=\frac{5}{2}$) atoms form corner sharing triangles, topologically equivalent to a regular kagom\'e lattice (Fig.~\ref{kagome}), with a strong Heisenberg character ($L \simeq 0$). 

\begin{figure}[ht]
\includegraphics[width=\columnwidth]{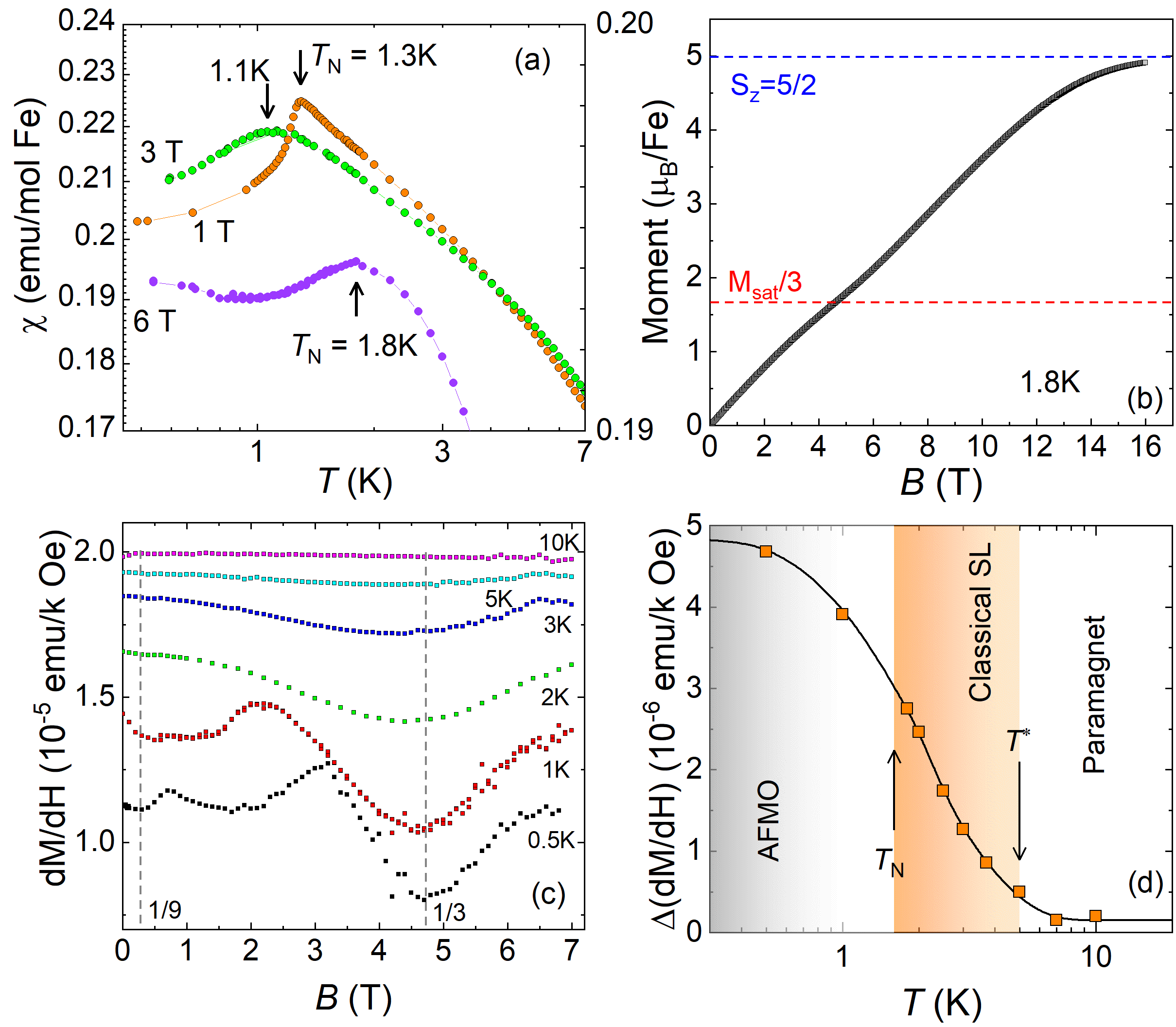}
\caption{\label{chi} (Color online) 
(a) Magnetic susceptibility $\chi$ measured under 1~T, 3~T (left axis) and 6~T (right axis) showing  antiferromagnetic transitions at $T_{\rm N}$.
(b) M vs $B$ measured at $T=1.8$~K shows a saturated value  $M_{\rm sat} \sim 5$~$\mu_B$/Fe and a 1/3-plateau for $B \sim 4.75$~T. 
(c) Derivative $dM/dH$ as a function of temperature (with a constant offset for clarity).  The dip at $\sim 5$~T signals the onset of the 1/3 plateau for $T=T^*$. 
(d) Evolution of the dip of the derivative of the magnetization, $\Delta (dM/dH)$, with temperature (orange squares). Line is guide to the eye. The classical spin liquid phase emerges below $T^* \simeq 5$~K, followed by an antiferromagnetic order (AFMO) below $T_{\rm N}$. 
}
\end{figure}

The macroscopic magnetic susceptibility is shown as a function of temperature in Fig.~\ref{chi}a. The susceptibility obeys a Curie-Weiss law, $C/(T-\theta_{\rm CW})$, with $\theta_{\rm CW} = -11(2)$~K, corresponding to an antiferromagnetic exchange interaction $J\sim 1$~K \cite{Supp}.
At $T=T_{\rm N}=1.3$~K a sharp peak in the magnetic susceptibility signals the onset of an antiferromagnetic order (AFMO) under 1~T (Fig.\ref{chi}a). The transition temperature $T_{\rm N}$ is reduced as compared to $|\theta_{\rm CW}| = 11$~K because of the frustration and the bidimensionality inherent to the kagom\'e geometry. Li$_9$Fe$_3$(P$_2$O$_7$)$_3$(PO$_4$)$_2$ thus shows a correlated paramagnetic regime between 1.3 and $\sim 10$~K. We also note that under a 6~T field the peak amplitude is greatly reduced and $T_{\rm N}$ slightly increases to 1.8~K, in agreement with NMR (see below), which suggests a non-trivial phase diagram.  

Above 10~K, $S=5/2$ spins are in a paramagnetic regime and M($B$) evolves almost linearly from 0 to 16~T. As $T$ is decreased, a pronounced curvature is observed towards the saturated value $M_{\rm sat}=5$~$\mu_B$, as shown in Fig.~\ref{chi}b. For $T<T^{*} \simeq 5$~K the slope of the magnetization clearly diminishes before increasing again, in the vicinity of $\sim M_{\rm sat}/3$. Fig.~\ref{chi}c shows the derivative $dM/dH$  which firmly confirms the existence of a narrow 1/3 magnetization plateau developing below $T^{*}$. The evolution of the 1/3 plateau is quantified by the size of the dip, $\Delta (dM/dH)$, and is shown as a function of $T$ in Fig.~\ref{chi}d. It displays a smooth crossover from a paramagnet for $T > T^{*}$ to the plateau phase, i.e. a classical collinear spin liquid phase formed by the set of the $uud$ states that remain degenerate at $M = M_{\rm sat}/3$, with no clear sign of symmetry-breaking like for a gas-liquid type transition. In the $T \ll |\theta_{\rm CW}|$ limit, its width can be estimated theoretically from Ref.\cite{Zhitomirsky2002}  to $\sim 1.5$~T  in agreement with our experimental value of $\sim 1.6$~T  (FWHM). 

The most recent theoretical phase diagram under applied fields has been computed using numerical tensor network methods. It shows three different incompressible phases or magnetization plateaus besides the 1/3 plateau\cite{Picot2016}. The $1/9$-plateau is only expected for the most quantum case ($S=\frac{1}{2}$). The others (one-magnon and two-magnon plateaus \cite{Picot2016}) should appear for all values of $S$ and lie very close to the saturation, respectively at 43/45 and 41/45 for $S=\frac{5}{2}$. These plateaus remain absent in our 1.8~K data but a clear conclusion is difficult and would require applying magnetic fields higher than 16~T. The absence of the $C_6$ rotational symmetry of our $S=\frac{5}{2}$ Fe$^{3+}$ lattice may also be detrimental to these semiclassical phases.

In order to gain more insights into the magnetic properties of LiFePO, we performed $^{31}$P NMR measurements. Li$_9$Fe$_3$(P$_2$O$_7$)$_3$(PO$_4$)$_2$ naturally contains two very sensitive NMR nuclei: $^{31}$P ($I=1/2$, $\gamma/ 2\pi = 17.23467$~MHz/T) and $^{7}$Li ($I=3/2$, $\gamma/ 2\pi = 16.54607$~MHz/T). The NMR spectra were obtained with $\nu_{0} = \gamma B_{0} / 2\pi =109.732$~MHz for the $^{31}$P local probe with $I=1/2$, avoiding any quadrupolar effects on the NMR resonance line, using the standard $\pi/2 - \tau - \pi$ pulse sequence, with $\tau = 20$~$\mu$s.

\begin{figure}[ht]
\includegraphics[width=\columnwidth]{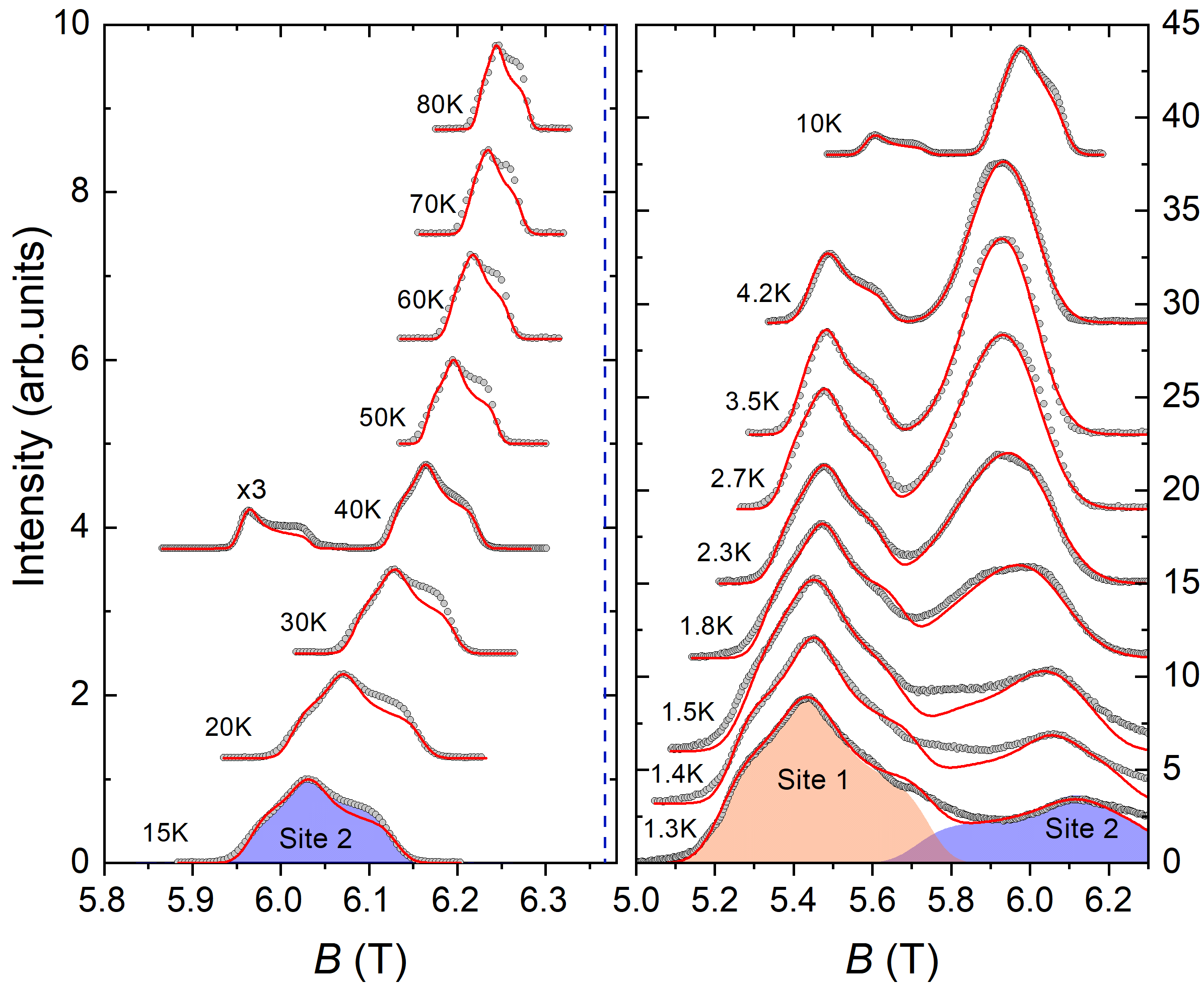}
\caption{\label{Fig_NMR} (Color online) 
Temperature dependence of the $^{31}$P NMR spectra (circles) measured as a function of applied magnetic field $B$ for LiFePO, from 80~K to 1.3~K. The phosphorus located on site (1) is only observed at low $T$ (magnified at 40~K ($\times 3$)), due to strong $T_2$ relaxation effects. Red line is a fit to Eq.\ref{Kshift} (see text) from which the tensor components $K$ are extracted.
Vertical dashed line indicates the reference field $B_{0}=6.3669$~T.  
}
\end{figure}

Fig.~\ref{Fig_NMR} displays the evolution of the $^{31}$P NMR spectrum with temperature. Two NMR lines, accounting for the two $^{31}$P sites (labelled as 1 and 2) expected from the structure \cite{Poisson1998} (Fig.\ref{kagome}a,b), are clearly visible below 10~K. The NMR intensity relative to site (1) gradually decreases as the temperature is increased, and finally escapes our NMR time window above 50~K, due to a fast relaxation, i.e. a shortening of the $T_2$ transverse relaxation time ($T_2 < 5-10$~$\mu$s).
The $^{31}$P NMR Knight shift $\textbf{K} = \pmb{A_{\rm hf} \chi } / \left( g\mu_B \hbar \gamma \right)$ provides a direct measurement of the local magnetic susceptibility tensor $\pmb{\chi}$. Following the common convention\cite{KS_convention} in spherical coordinates, the eigenvalues of $\textbf{K}$ relate to the shift of the NMR line $\Delta B = B_0 - B$ in the applied field $B$ through \cite{Slichter}:
\begin{equation}
\label{Kshift}
\frac{\Delta B}{B} = K_{\rm iso} + K_{\rm ax}\left( 3\cos^2 \theta -1 \right) + K_{\rm ani}\sin^2 \theta \cos 2\phi\\
\end{equation}
%

\begin{figure}[ht]
\includegraphics[width=\columnwidth]{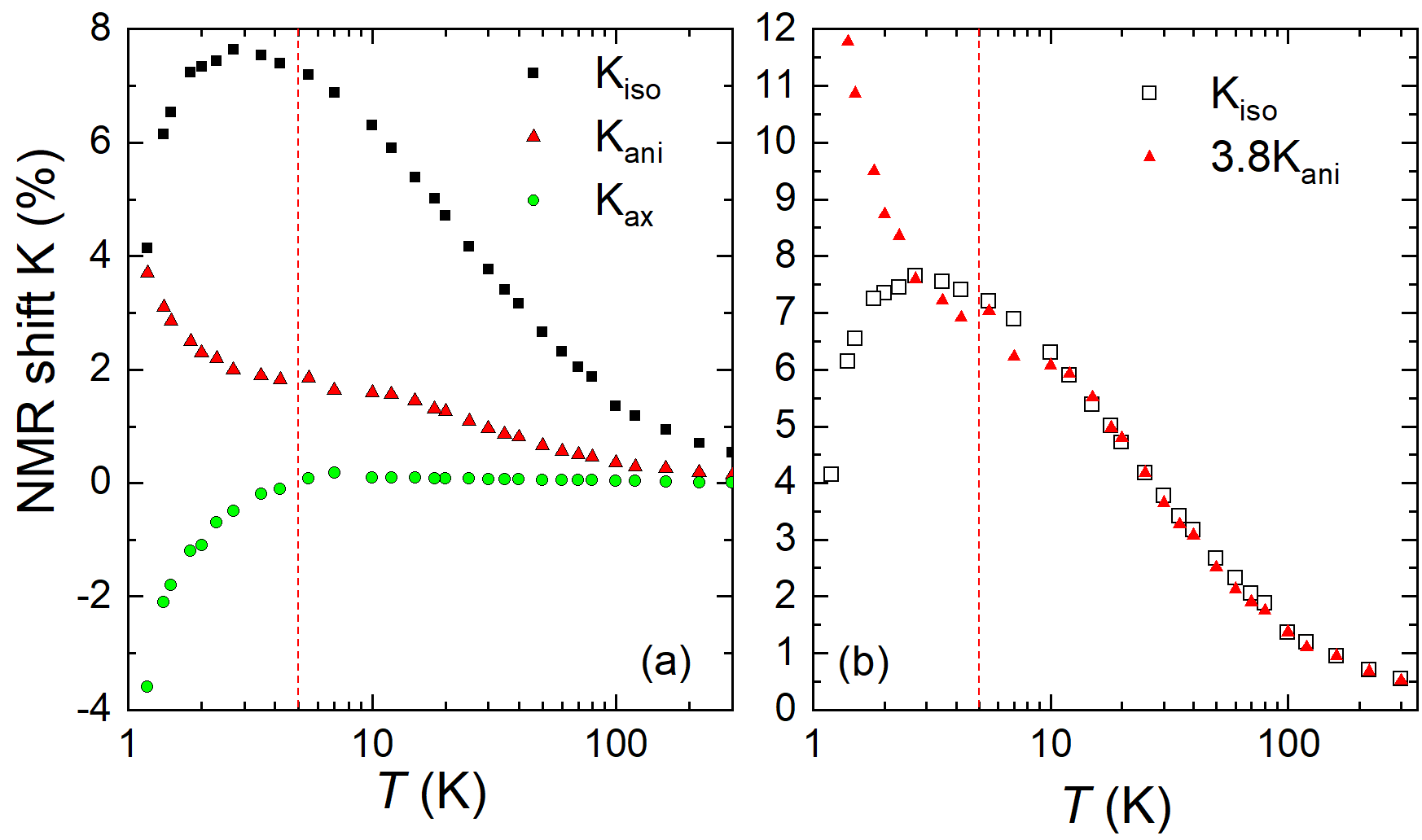}
\caption{\label{Fig_K} (Color online) (a) Components of the NMR shift \textbf{K} as a function of temperature for site 2.  (b) In the paramagnetic regime, $K_{\rm iso}$ and $K_{\rm ani}$ scale with each other showing that the magnetic susceptibility is isotropic above $T^*$. The red vertical dashed line indicates $T^*$. }
\end{figure}

The spectra for sites (1) and (2) could be well fitted using Eq.\ref{Kshift} for an isotropic powder distribution (Fig.~\ref{Fig_NMR}). The mismatch between observed (circles) and simulated (red line) intensities is due to an anisotropic $T_2$ spin-spin relaxation time, as confirmed by measurements using different duration $\tau$ between the pulses (not shown). While site (1) and site (2) probe the same magnetic Fe$^{3+}$ ions, the anisotropic components of the Knight shift tensor are best revealed from site (2) and we thus focus on that NMR line in the following \cite{Supp}. We then extract the temperature dependence of the components of the Knight shift tensor for site (2), as shown in Fig.~\ref{Fig_K}. 

In the paramagnetic regime ($300-10$~K), the NMR lineshape of site (2) is anisotropic and shows a pronounced shoulder on its right-hand side. However, above 10~K, $K_{\rm iso}$ and $K_{\rm ani}$ are found to scale with each other (Fig.~\ref{Fig_K}) and thus the spectrum could be well fitted using a unique $T$-dependent parameter --an isotropic magnetic susceptibility $\chi(T)$--, with $K_{\rm iso}= A_{\rm hf}^{\rm iso} \chi(T)$, $K_{\rm ani}= A_{\rm hf}^{\rm ani} \chi(T)$ and $K_{\rm ax} \simeq 0$. The $T$-independent  components of the hyperfine tensor per Fe atom are $A_{\rm hf}^{\rm iso} =2.21(3)$~kOe$/\mu_B$ and $A_{\rm hf}^{\rm ani} =0.591(5)$~kOe$/\mu_B$ for site (2).
An anisotropic hyperfine coupling tensor is a usual feature of  $^{31}$P NMR. 

Upon cooling below $T^*=5$~K the NMR lineshape evolves and reveals the development of a strongly temperature-dependent anisotropy, which can thus only be ascribed to the magnetic susceptibility. The decrease of $K_{\rm iso}$ signals the development of antiferromagnetic correlations when $T<|\theta_{\rm CW}| = 11$~K, and clearly indicates the onset of a correlated magnetic phase. Both site (1) and site (2) reveal an anisotropic susceptibility as $K_{\rm ani}$ and $K_{\rm ax}$ undergo an abrupt change, a remarkable feature for Heisenberg spins. The correlated magnetic phase observed for $B \sim 6$~T and $1.3 < T < 10$~K, for which $M \sim \frac{1}{3}M_{\rm sat}$,  seems thus consistent with the theoretically expected, strongly anisotropic,  collinear phase.
%

\begin{figure}[ht]
\center
\includegraphics[width=0.8\columnwidth]{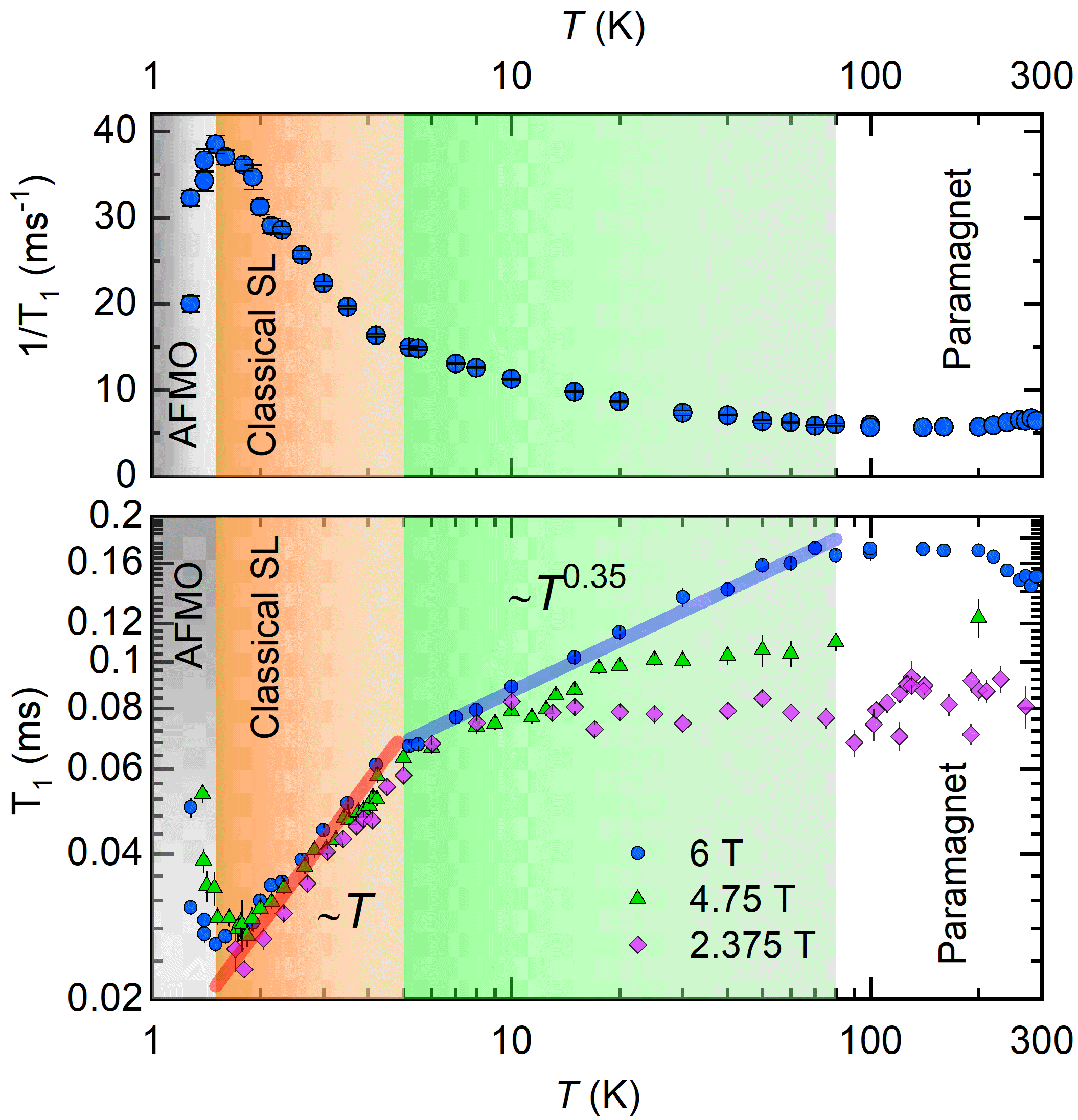}
\caption{\label{Fig_T1} (Color online) $1/T_1$ (up) and $T_1$ (bottom) probed by $^{31}$P NMR as a function of temperature in LiFePO. Four different regimes can be identified: a static antiferromagnetic-type order (AFMO, grey), a classical spin liquid phase (orange), an atypical high-$T$ regime (green) and high-$T$ paramagnetic phase. The classical spin liquid phase displays spin dynamics with a linear law $T_1 = \beta T$ (red line, see text).}
\end{figure}

We now use $^{31}$P NMR to investigate the magnetic fluctuations on and around this 1/3-plateau phase. The spin-lattice relaxation rate $1/T_1$ probes the imaginary part of the dynamic susceptibility $\chi''_{\perp}(\nu,q)$ at low energy ($\nu \sim 100$~MHz) through:
\begin{equation}
\label{T1}
\frac{1}{T_1} = \frac{\gamma^2}{g^2 \mu_B^2}k_B T \sum_q \left| A_{\rm hf}(q) \right|^2 \frac{\chi''_{\perp}(\nu,q)}{2\pi \nu}  \\ 
\end{equation}
and was determined through the saturation recovery method, with a $\pi/2 - \Delta t - \pi/2  - \tau - \pi$ pulse sequence, from 1.3 to 300~K, at the maximal intensity of site (2) (Fig.~\ref{Fig_T1}) for three different applied fields (6.37~T, 4.75~T and 2.375~T)~\cite{Supp}. Four different regimes in temperature can be identified. A peak is clearly observed for $T \sim 1.5$~K, close to the transition temperature $T_{\rm N}$, which separates the static magnetic order from a classical spin-liquid state for $T_{\rm N} < T < T^*$ for which $T_1 \sim T$ (thick red line in Fig.~\ref{Fig_T1}). An estimate of the real exponent is difficult for such narrow regime, notably because of critical fluctuations expected just above $T_{\rm N}$.
A crossover is observed at $T=T^*\sim 5$~K, and the dynamics is now dominated by a higher spin-lattice relaxation rate for $T^*<T<80$~K, with an atypical power law behavior $\alpha = 0.35(3)$ (blue line) observed for $B = 6.37$~T, which persists well above $|\theta_{\rm CW}|$, up to $\sim 80$~K. For this regime, the relaxation is clearly impacted by the applied field, and we speculate that lattice vibrational modes\cite{Vega2006} are responsible for the unusual  behavior observed here. 
Finally, $T_1$ levels off in the high-$T$ paramagnetic regime. The Moriya paramagnetic limit gives the $T$-independent rate $1/T_1^{\infty} = \gamma^2 g^2 A_{\rm hf}^2S(S+1)/3z_1 \sqrt{2\pi} \nu_e$~\cite{Moriya1956,Kermarrec2014} in magnetic insulators, where $\nu_e = J \sqrt{2zS(S+1)/3} / h$ the exchange frequency, with $z_1=2$ the number of probed Fe and $z=4$ the number of magnetic nearest-neighbours of a Fe atom. Using $A_{\rm hf} = A_{\rm hf}^{\rm iso}$ and the experimental value $T_1^{\infty} = 0.05(3)$~ms extrapolated at $B \rightarrow 0$~\cite{Supp}, the formula leads to $J = 0.7(3)$~K, in reasonable agreement with $J\sim 1$~K. We note that lithium diffusion likely starts to contribute to the relaxation above 250~K. 

We now discuss more specifically the classical spin liquid phase observed in the $T$-range $1.5-5$~K in the context of theoretical predictions. The low$-T$ dynamics of the classical Heisenberg kagom\'e antiferromagnet has been investigated early on by Monte-Carlo simulations \cite{Chalker1992,Keren1994,Reimers1993}. They gave evidence for the role of thermal fluctuations in selecting a coplanar short-range order via the order-by-disorder mechanism for $T \rightarrow 0$. Short-range nematic spin correlations are predicted to survive at low $T$ with non-dispersive soft magnon modes\cite{Harris1992}, a characteristic feature associated with the large degeneracy inherent to the kagom\'e lattice. Numerical simulations of the spin autocorrelation function yield $\left\langle \bm{S}(0) \bm{S}(t) \right\rangle \sim e^{- \nu t}$  for $H=0$ \cite{IFM, Keren1994,Robert2008, Taillefumier2014}, with $\nu = ck_B T / \hbar S$ ($c \sim 1$)\cite{IFM, Moessner1998}. The kagom\'e bilayer compound SrCr$_8$Ga$_4$O$_{19}$ was the first experimental candidate to hint at such spin dynamics, yet with an important contribution from quantum fluctuations \cite{Uemura1994,Keren1994,BonoPRL2004,Lee1996} and spin vacancies \cite{Limot2002}. Later on, inelastic neutron scattering exposed a $T$-linear inverse relaxation rate in deuterium jarosite over the extended temperature range $0.06 < T/J_{\rm cl} < 1$ ~\cite{FakEPL2008}, with $J_{\rm cl} = JS(S+1)$. Here, the $T$-dependence of $1/T_1$ can be inferred from $\nu$ through the Redfield formula:
\begin{equation}
\label{Red}
\frac{1}{T_1} = \frac{2 \gamma^2 \Delta^2 \nu}{\nu^2 + \gamma^2 B^2} \simeq \frac{2 \gamma^2 \Delta^2 }{\nu} \\ 
\end{equation}
where $\Delta$ is the amplitude of the fluctuating field at the phosphorus position. Using the constant high-$T$ value $T_1^{\infty}$, this  leads to $T_1  = \beta T$ with $\beta= ck_B T_1^{\infty} / \nu_e \hbar S$. In LiFePO, the dynamical behavior of $T_1$ could be well fitted to such a law, with no adjustable parameter but $c = 1.2(1)$ (thick red line in Fig.~\ref{Fig_T1}), giving firm evidence of the spin liquid phase within the $T$-range $0.2<T/J_{\rm cl}<0.6$ with $J_{\rm cl}  = 8.25$~K.

In a more recent work\cite{Taillefumier2014}, Taillefumier \textit{et al.} characterized the dynamics for the long time scales appropriate for NMR using semiclassical numerical techniques, and predicted two different low temperature regimes in the $B\rightarrow 0$ limit: (\textit{i}) a cooperative, classical, spin-liquid regime with $T_1 \sim T$ in line with what is discussed above and (\textit{ii}) a coplanar dynamical regime for $T/J_{\rm cl}<0.005$ with a strong anisotropic relaxation dominated by weathervane defects modes. Since $J_{\rm cl}  = 8.25$~K, our measurements down to 1.3~K in LiFePO give only access to the region $T/J_{\rm cl} > 0.2$, i.e. the high-$T$ paramagnetic and cooperative spin liquid phases, before residual interactions apparently lift the degeneracy and produce the antiferromagnetic  transition. The observed thermal behavior of our spin-lattice relaxation rate, $T_1 \sim T$, seems to show that this dynamics unexpectedly persists under moderate applied fields ($B \leq 6$~T).  This can perhaps be understood because similar zero-energy excitations exist within the $uud$ manifold of the 1/3 plateau. Furthermore, the short-range anisotropic correlations that develop above $T_N$ according to our NMR local susceptibility measurements may indicate either that the nematic correlations are still present for $T_N < T < T^*$, or that we lie close to a quasi-static magnetic order. In the light of our results, a theoretical study of the spin dynamics under applied fields, in and around the 1/3-phase, would be highly relevant to confirm the origin of the dynamics observed in NMR.

%
In conclusion we have identified a new classical $S=\frac{5}{2}$ Heisenberg kagom\'e antiferromagnet that displays the 1/3 magnetization plateau predicted a long time ago. 
Our $^{31}$P NMR measurements further identified a classical spin-liquid regime for $T_N < T < T^*$, for which the time scale of the spin dynamics is set by the temperature only, with $T_1 \sim T$, in very good agreement with numerical estimates.
Additional experimental and theoretical work would be needed to single out the true signatures of the nematic spin liquid of the 1/3 plateau from the spin fluctuations inherent to the kagom\'e lattice, and how the latter evolve under large applied magnetic fields.
%
The discovery of archetypal magnetic frustration relevant to the kagom\'e lattice in LiFePO offers great perspectives to improve our knowledge on classical and quantum fluctuations effects in zero and applied fields in the KHAF. In LiFePO, further elastic and inelastic neutron scattering studies could help revisiting the central question of the selection of the $q=0$ or $q=\sqrt{3} \times \sqrt{3}$ ground-state in such model and also confirming the relevance of a simple, nearest-neighbour exchange Hamiltonian. LiFePO and other members of this family [Cr$^{3+}$ ($S=\frac{3}{2}$), V$^{3+}$ ($S=1$)] will certainly open new avenues to search for quantum phases relevant to the KHAF.

\section*{Acknowledgments}
It is a pleasure to acknowledge useful discussion with Prof. S. Greenbaum on lithium diffusion, P. Berthet and C. Decorse for sharing their expertise on samples synthesis of lithium oxides. EK and RS acknowledge financial support from the labex PALM for the QuantumPyroMan project (ANR-10-LABX-0039-PALM). BKR thanks DST for providing Inspire faculty award and S. Srinath (UoH) for allowing the sample preperation. This work was supported by the French Agence Nationale de la Recherche under grant ANR-18-CE30-0022-04 ‘LINK’.


\begin{thebibliography}{99}
\bibitem{KS_convention}
If $(x,y,z)$ is the vector basis in which the Knight shift tensor $\textbf{K}$ is diagonal, then the components used here are defined as $K_{\rm iso} = \frac{1}{3}\left( K_{xx} + K_{yy} + K_{zz} \right)$, $K_{\rm ax} = \frac{1}{6}\left( 2K_{zz} - K_{xx} - K_{yy} \right)$ and $K_{\rm ani} =\frac{1}{2}\left( K_{yy} - K_{xx}\right)$
\end{thebibliography}
\end{document}